\documentclass[aip, rsi, reprint, citeautoscript]{revtex4-1}
\usepackage{graphicx}
\usepackage{amsmath}
\usepackage{amssymb}
\usepackage{xcolor}
\usepackage{ulem}

\draft % marks overfull lines with a black rule on the right
\begin{document}

\title{Systematic investigation of wear-induced cold welding in ultrahigh vacuum piezoelectric motors with non-metallic coatings} %Title of paper

\author{Taemin Ahn}
\affiliation{Department of Physics, Pohang University of Science and Technology (POSTECH), Pohang 37673, South Korea}

\author{Sungmin Song}
\affiliation{Department of Physics, Pohang University of Science and Technology (POSTECH), Pohang 37673, South Korea}

\author{Ungdon Ham}
\affiliation{Center for Artificial Low Dimensional Electronic Systems, Institute for Basic Science (IBS), Pohang 37673, South Korea}

\author{Tae-Hwan Kim}
\email[Author to whom correspondence should be addressed: ]{taehwan@postech.ac.kr}
\affiliation{Department of Physics, Pohang University of Science and Technology (POSTECH), Pohang 37673, South Korea}

\date{\today}

\begin{abstract}
Piezoelectric motors are widely used in various applications where both precision positioning and miniaturization are required.
Inertial or quasi-static motors are commonly employed because of their high accuracy, which demands 
consistent sliding friction between moving sliders and their static counterparts for reliable operation. 
In general, slider wear is unavoidable after long-term use.
This wear can often lead to more serious cold welding in vacuum, which is also referred to as friction welding induced by direct contact between similar metal surfaces.
Non-metallic coatings can prevent such unwanted cold welding in ultrahigh vacuum (UHV) applications.
However, the practical reliability of available coatings under UHV conditions still remains to be elucidated.
Here, we systematically investigate the practical reliability of commonly used UHV-compatible lubricant coatings for piezoelectric motors in vacuum.
We demonstrate that polytetrafluoroethylene (PTFE) shows the most reliable long-term operation in vacuum, while other coatings eventually lead to wear-induced cold welding and motor failure.
Our findings provide a simple and effective method to improve the long-term performance of UHV piezoelectric motors by coating the
slider surface with PTFE.

\end{abstract}


\maketitle %\maketitle must follow title, authors, abstract and \pacs

\section{Introduction}

Piezoelectric motors, which have long travel ranges with high positioning accuracy, are widely used in various fields where nanopositioning with small physical dimensions is required~\cite{ru_nanopositioning_2016}.
Inertial or quasi-static piezoelectric motors show reliable and accurate single-step motions~\cite{ouyang_micro-motion_2008}.
Such precise nanopositioning can only be achieved if the kinetic friction between moving sliders and their static counterparts is kept constant~\cite{persson_sliding_2011}.

In general, unwanted but unavoidable slider wear due to the friction becomes a major obstacle to the long-term operation of motors~\cite{wallaschek_contact_1998, ru_nanopositioning_2016, puoza2021UltrasonicMotorsStructurala}.
Especially, in vacuum environments, the lack of proper lubrication between the sliders and their static counterparts results in higher friction compared to ambient conditions.
As a result, oxide layers on the metal surfaces of moving parts can easily peel off, leading to wear-induced cold welding through the diffusion of metal atoms~\cite{Yates2015, merstallinger2009AssessmentColdWelding, pashley_adhesion_1984,lu_cold_2010, budakian_time_2002}.

It is particularly important to avoid cold welding in piezoelectric motors, which are widely used in ultrahigh vacuum (UHV) scanning probe microscopes (SPMs). % to investigate atomic-scale surface states.
To prevent wear-driven cold welding in vacuum applications, non-metallic robust sliders~\cite{wolkow_variable_1992, wittneven_low-temperature_1997, guo_note_2014, pan_3he_1999, den_heijer_improving_2014,  wu_low-temperature_2018} or dry lubricants on sliders~\cite{fernandez-lomana_millikelvin_2021} have been chosen.
However, the possibility of the failure of the non-metallic sliders still exists after long-term operation in UHV due to changes in friction between the slider and its static counterparts~\cite{wu_low-temperature_2018}.
To overcome this problem~\cite{wu_low-temperature_2018, liu_new_2018,den_heijer_improving_2014}, the complicated \textit{in-situ} friction adjustment is inevitable.
Such intricacy makes it very difficult to design ultra-compact piezoelectric motors, which are ideal for low-temperature SPMs.
On the other hand, metal sliders have often been used in simple and compact SPMs~\cite{cherepanov_nanopositioner_2012}.
In this case, the reliable motion of the sliders is even more critical after long-term operation in vacuum, because it is almost impossible to adjust the friction without breaking vacuum due to their simple structure.
Therefore, it is necessary to find suitable coatings for metal sliders for their reliable long-term operation in vacuum.

Here, we systematically investigate various UHV-compatible coatings on metal sliders for the practical long-term reliability of a piezoelectric motor under vacuum.
Long-term operation for each coating is performed automatically using a vision-based positioning method.
Our results reveal that a polytetrafluoroethylene (PTFE) coating exhibits the most consistent long-term operation in the UHV environment.
In sharp contrast, other coatings become stuck after some operations in vacuum. 
This finding can be directly applied to other UHV piezoelectric motors without significant modification.

\section{Experimental setup}
\begin{figure}[t]
\includegraphics[width=0.9\linewidth]{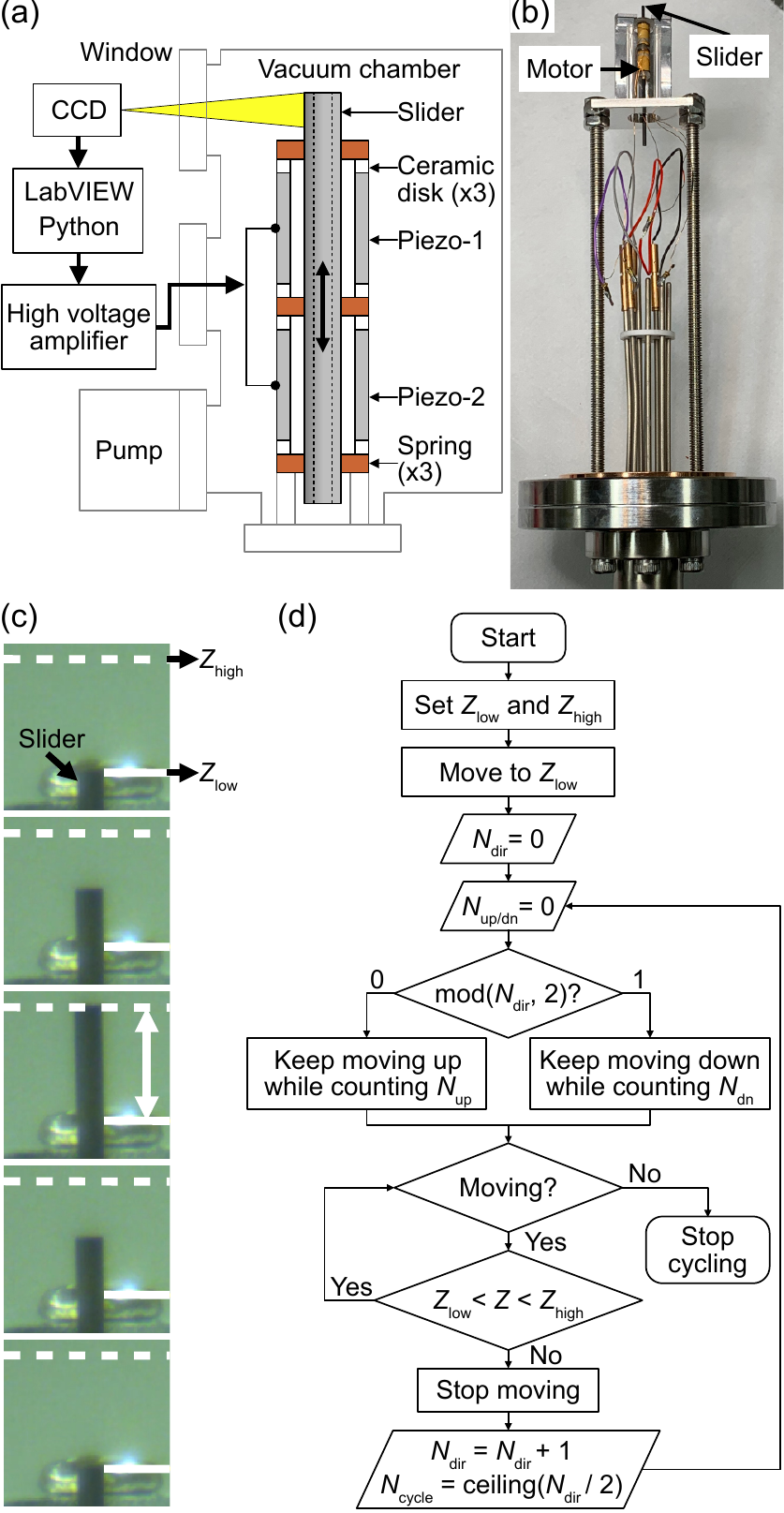}
\caption{\label{fig01}
(a) Schematic diagram of the experimental setup used to test the KoalaDrive-type piezoelectric motor in vacuum. 
(b) Photo of a homemade holder used to secure the piezoelectric motor in vacuum. 
(c) Snapshots capturing the motion of the slider in  vacuum. 
(d) Flowchart of automated reliability tests performed on each coated slider.
Here, $Z_{\rm low}$ and $Z_{\rm high}$ represent the lowest and highest positions within each cycle.
$N_{\rm up/dn}$ and $N_{\rm cycle}$ denote the cumulative number of up/down steps per cycle and the total number of cycles during the test, respectively.
}
\end{figure}

In this work, a homemade KoalaDrive-type quasi-static piezoelectric motor~\cite{wang2009simple, cherepanov_nanopositioner_2012, lu2012comment} is used in a vacuum chamber [Fig.~\ref{fig01}(a)].
The motor consists of a metal slider, three BeCu springs (TE Connectivity, part number: 5050865-6), three ceramic disks, and two tube piezo elements (EBL Products, Inc.; part number: EBL \#2; outer diameter: 3.2~mm; wall thickness: 0.25~mm; length: 6.3~mm).
The metal slider is made of a stainless steel (SS) tube (Needle Store; part number: OD-10-032; outer diameter: 0.70~mm; wall thickness: 0.15~mm; length: 35~mm).
To test long-term operation, we set a bipolar driving voltage of $\pm100$~V.
This voltage is much higher than a typical threshold voltage ($\pm40$~V) below which the motor does not move with a pristine SS tube.

We use a homemade holder [Fig.~\ref{fig01}(b)] to mount the piezoelectric motor, which was attached to the vacuum chamber. 
The motion of the slider is monitored through a window by a CCD camera placed outside the vacuum chamber [Fig.~\ref{fig01}(c)], and the position of the slider was identified by RGB color difference between the background and the slider from the CCD image. 
To automate the long-term operation of the piezoelectric motor without human intervention, we have developed our own Python and LabVIEW programs. 
After each long-term operation test in vacuum, we ventilated the vacuum chamber before replacing the slider of the piezoelectric motor. 
After replacing another slider, we pumped down the chamber ($P<1\times10^{-8}$~Torr) for the next long-term operation test with the replaced slider.

\begin{figure}[t]
\includegraphics[width=0.9\linewidth]{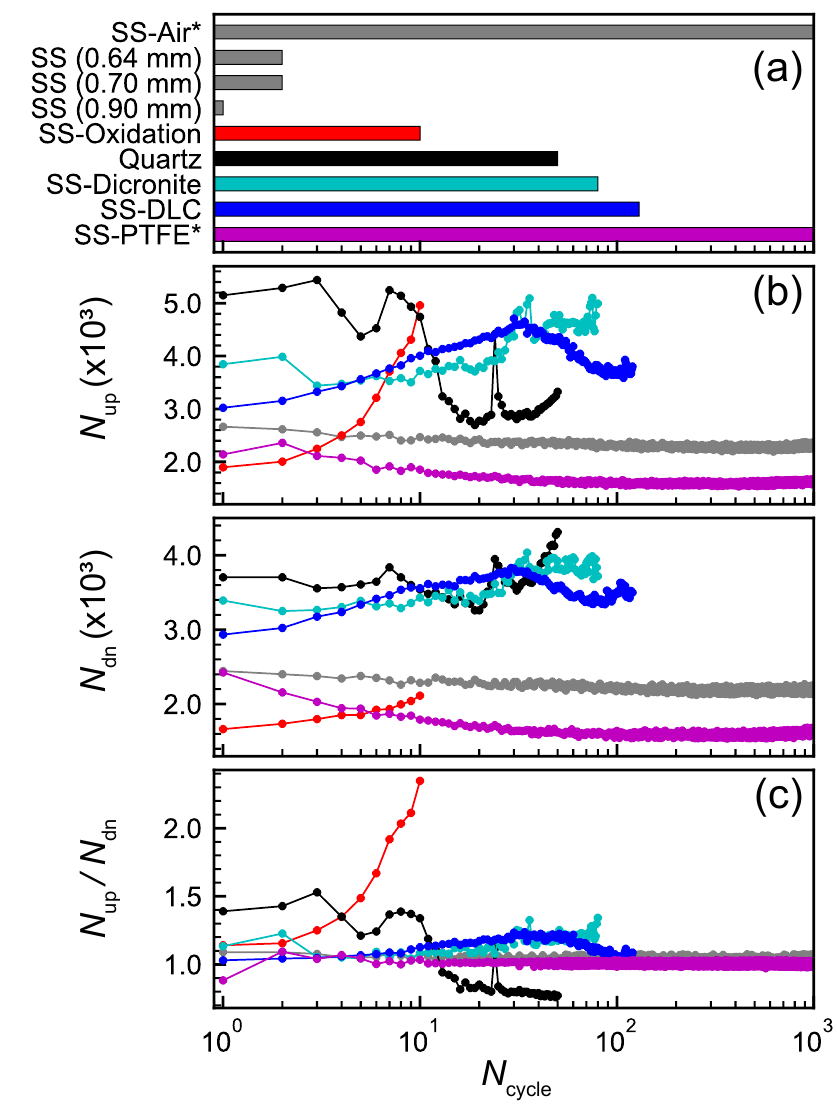}
\caption{\label{fig02}
(a) Comparison chart showing the cumulative cycles of bare stainless steel (SS) tubes and a quartz tube, as well as coated SS tubes.
SS tubes with a diameter of 0.70~mm are coated with four different coatings [thick oxide, Dicronite\textsuperscript{\textregistered}, diamond-like carbon (DLC), and polytetrafluoroethylene (PTFE)],
while bare SS tubes with three different diameters of 0.64, 0.70, and 0.90~mm are tested.
All tests were performed under vacuum until forced to stop, except for a PTFE coated SS tube (SS-PTFE*).
For comparison, we have also included the test result of a bare SS tube in air (SS-Air*), which has similar characteristics to the PTFE coated tube.
(b,c) Comparison plots of (b) cumulative steps per cycle during up ($N_{\rm{up}}$) and down ($N_{\rm{dn}}$) motions, and (c) the ratio of up to down steps ($N_{\rm{up}}/N_{\rm{dn}}$).
}
\end{figure}

Figure~\ref{fig01}(d) shows a detailed flowchart of the automated reliability test.
Before the first test run, we initialize the start position $Z_{\rm{low}}$ and end position $Z_{\rm{high}}$ from the CCD images to ensure the same travel length ($\sim3$~mm) during each test run.
Once the slider is automatically positioned at $Z_{\rm{low}}$ by the CCD camera, all variables used are set to zero.
Then, the slider starts moving up (down) when $N_{\rm{dir}}$ is even (odd).
$N_{\rm{up/dn}}$ and $N_{\rm{cycle}}$ indicate the cumulative up/down steps per cycle and the total number of cycles during the test, respectively.
When the slider reaches $Z_{\rm{high}}$, the controller automatically reverses the moving direction of the slider.
Until stopped by increasing friction, a cycle of up and down motion will continue.

\section{Results and discussion}
Subsequently, we investigate $N_{\rm{cycle}}$ for different types of coatings on the slider.
A bare SS tube (diameter: 0.70~mm), which worked well in air, becomes stuck after only two cycles in vacuum [see SS-Air* and SS (0.70~mm) in Fig.~\ref{fig02}(a)].
In contrast, the bare SS tube moves again without any treatment after air ventilation.
We attribute the surprising recovery of motion to the regenerated oxide layer on the SS tube after exposure to air~\cite{merstallinger2009AssessmentColdWelding}.
During the vacuum test, the pristine oxide layer on the SS tube is broken by wear due to the friction between the BeCu springs and the SS tube.
The damaged oxide layer is quickly restored in air, but gets worse with time in vacuum, resulting in the inevitable cold welding~\cite{podgornik_performance_2012}.

In addition, we reduce friction and wear in the SS-BeCu contacts by replacing them with a thinner SS tube (diameter: 0.64~mm) to see any improvement in vacuum.
While the thinner SS tube slightly lowers the threshold voltage for motion in air, it also becomes stuck after several cycles in vacuum [see SS (0.64~mm) in Fig.~\ref{fig02}(a)].
We experimentally confirm that our specific setup shows reliable motion in air with bare SS tubes for the diameter range of 0.64--0.90~mm.
Beyond this diameter range, the motor did not work properly even in air.
In this sense, we choose SS tubes with a diameter of 0.70~mm as a reference to evaluate the relative efficiency of the coating materials in this work.

To obtain thicker oxide layers against the wear in vacuum, we further oxidize the SS tube by heating it to $\approx500$~K in air for 2~h~\cite{reghuraj2017black, ferreira2001influence}.
However, the oxidized SS tube is forced to stop after only 10 cycles [red bar in Fig.~\ref{fig02}(a)].
Before the motor gets stuck, $N_{\rm{up/dn}}$ increases significantly with each cycle [red dots in Figs.~\ref{fig02}(b)].
Furthermore, the ratio of $N_{\rm{up}}$ to $N_{\rm{dn}}$ ($N_{\rm{up}}/N_{\rm{dn}}$) also increases with each cycle [red dots in Fig.~\ref{fig02}(c)].
We can conclude that the even thicker oxide layer is not enough to ensure  reliable long-term operation in vacuum.

To prevent cold welding due to direct contact between the SS tube and the BeCu springs, we can replace the SS tube with a non-metallic quartz tube of the same dimensions.
We then examine the long-term reliability of the quartz tube in the same way as before.
After about 50 cycles [black bar in Fig.~\ref{fig02}(a)], the quartz tube gets stuck with the fluctuating $N_{\rm{up}}$/$N_{\rm{dn}}$ [black dots in Figs.~\ref{fig02}(b) and \ref{fig02}(c)].
Although the quartz tube shows better performance compared to the bare or oxidized SS tube, 
the observation strongly suggests that the kinetic friction increases significantly with each cycle before being forced to stop.
Furthermore, the quartz tube does not move even after air ventilation, in sharp contrast to the SS tube.
Note that we repeat the same test with a new quartz tube in air, resulting in more or less the same $N_{\rm{cycle}}$.
We therefore exclude the quartz tube as a candidate for the reliable sliders in our piezoelectric motor.

Alternatively, we try to coat UHV-compatible dry lubricants such as Dicronite\textsuperscript{\textregistered}\cite{Dicronite} (coating thickness: $\le 0.5$~$\mu$m) and diamond-like carbon (DLC)\cite{vanhulsel2007dlc} (coating thickness: $\sim$1~$\mu$m) on the bare SS tube to reduce friction between the SS tube and the BeCu springs.
Both Dicronite\textsuperscript{\textregistered} and DLC coatings show more reliable results than the others discussed above [cyan and blue dots, respectively, in Figs.~\ref{fig02}(b) and \ref{fig02}(c)].
However, wear of the coatings is unavoidable.
The Dicronite\textsuperscript{\textregistered} and DLC coated SS tubes become stuck after about 80 and 130 cycles, respectively [cyan and blue bars in Fig.~\ref{fig02}(a)].
Note that there is no inherent up-down asymmetry in the case of a bare SS tube in air [gray dots in Fig.~\ref{fig02}(c)].
The observed up-down asymmetry in the Dicronite\textsuperscript{\textregistered} and DLC coatings can be attributed to uneven wear caused by imperfect alignment between the three sets of BeCu springs.

\begin{figure}[t]
\includegraphics[width=0.9\linewidth]{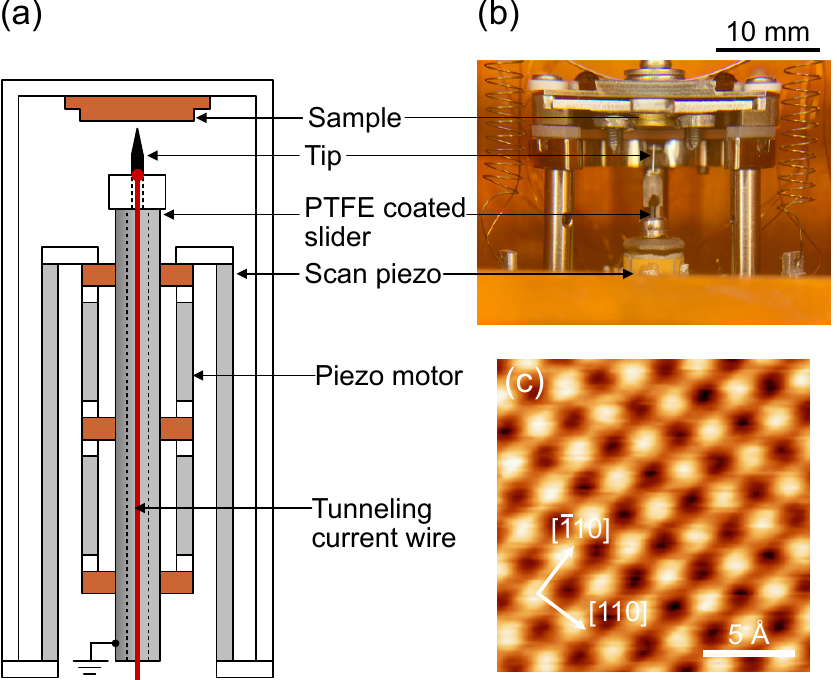}
\caption{\label{fig03}
(a) Schematic of a scanning tunneling microscope (STM) using a piezoelectric motor with a PTFE coated SS tube. 
The tunneling current wire is shielded by an electrically grounded SS tube.
(b) Photo of a homemade STM.
(c) Atom-resolved STM image of clean Cu(001) at 80~K. 
Imaging conditions: $V_{\rm b}=-3.0$~mV, $I_{\rm t}=134$~nA.
}
\end{figure}

PTFE coatings have also been widely used as UHV-compatible dry lubricants because of their low outgassing rate~\cite{nasa} and low friction coefficient under vacuum~\cite{yuan_study_2010}.
Unlikely other vacuum-compatible coatings, the PTFE~\cite{PTFE} coated SS tube (coating thickness: $\sim$10~$\mu$m) shows the best performance without seizing even after more than 1000 cycles [magenta bar in Fig.~\ref{fig02}(a)].
Although the PTFE coating peels off during the test cycles, the peel-off debris from the PTFE coating acts as an additional lubricant~\cite{demas_tribological_2008}.
As a result, the $N_{\rm{up/dn}}$ of the PTFE coated SS tube decreases with each cycle and remains almost unchanged since $N_{\rm{up/dn}}\approx 100$.
Furthermore, $N_{\rm{up}}$/$N_{\rm{dn}}$ reaches $\sim1$ after 100 cycles [magenta dots in Figs.~\ref{fig02}(b) and \ref{fig02}(c)].
Since we used identical BeCu springs and SS tubes for all tests, increasing the coating thickness resulted in a corresponding increase in frictional load. 
Interestingly, the thickest PTFE coating showed the best wear resistance despite the higher frictional load. 
This finding highlights the importance of the self-lubricating nature of PTFE for reliable operation in vacuum piezoelectric motors~\cite{demas_tribological_2008, rodiouchkina2021TribologicalBehaviourTransfer}.
Thus, we strongly suggest that the PTFE coated SS tube is the optimal slider for use in vacuum.

Finally, we demonstrate the performance of our piezoelectric motor by conducting STM experiments.
We built an STM combined with the piezoelectric motor using the PTFE coated slider.
The schematic drawing and photo of our homemade STM head with the piezoelectric motor are shown in Figs.~\ref{fig03}(a) and \ref{fig03}(b), respectively.
The piezoelectric motor is placed inside a scan piezo tube, which can be found elsewhere~\cite{cherepanov_nanopositioner_2012}.
The PTFE coated SS tube is also used as an electrical shield to minimize electrical noise in tunneling currents.
Thanks to our reliable piezoelectric motor, we successfully approach and obtain atom-resolved STM images of clean Cu(001) (prepared by several cycles of Ne$^+$ sputtering and annealing at 820~K) under UHV ($P<1.0\times10^{-10}$~Torr) at 80~K [Fig.~\ref{fig03}(c)].

\section{Conclusion}
In this work, we explored the long-term reliability of the piezoelectric motor under vacuum with various UHV-compatible coatings on the slider.
In particular, the PTFE coated slider shows the most reliable performance compared to the other dry lubricants tested, leading to the consistent motion with each cycle.
In addition, our homemade STM head equipped with the PTFE coated slider demonstrates successful low-temperature operation under UHV without any noticeable wear-induced cold welding.
We expect that the PTFE coating can be directly applied to other piezoelectric motors in various vacuum SPMs to avoid wear and cold welding.

\begin{acknowledgments}
% Put your acknowledgments here.
This work was supported by the National Research Foundation of Korea (NRF) funded by the Ministry of Science and ICT, South Korea (Grants No. NRF-2021R1F1A1063263, 2021R1A6A1A10042944, and 2022M3H4A1A04074153).
\end{acknowledgments}



\begin{thebibliography}{31}
\bibitem [{PTF()}]{PTFE}%
  \BibitemOpen
  \href@noop {} {}\bibinfo {note} {In our case, electrostatic spraying is
  employed for applying PTFE coating. Subsequently, the coated surface is cured
  by heating in order to bond the PTFE to the substrate. The PTFE coating on
  the surface of our stainless steel tubes was conducted by a Korean coating
  company (SUNG MIN TF Co., Ltd.).}\BibitemShut {Stop}%
\end{thebibliography}
\end{document}